# Topological phenomena demonstrated in photorefractive photonic lattices

Shiqi Xia[1], Daohong Song[1,2,5], Nan Wang[1], Xiuying Liu[1], Jina Ma[1], Liqin Tang[1,2], Hrvoje Buljan[1,3,6], and Zhigang Chen[1,2,4]

[1] *The MOE Key Laboratory of Weak-Light Nonlinear Photonics, TEDA Applied Physics Institute and School of Physics, Nankai University, Tianjin 300457, China*
[2] *Collaborative Innovation Center of Extreme Optics, Shanxi University, Taiyuan, Shanxi 030006, People's Republic of China*
[3] *Department of Physics, Faculty of Science, University of Zagreb, Bijenička c. 32, 10000 Zagreb, Croatia*
[4] *Department of Physics and Astronomy, San Francisco State University, San Francisco, California 94132, USA*
[5] *songdaohong@nankai.edu.cn*,   [6] *hbuljan@phy.hr*
*Corresponding author: zgchen@nankai.edu.cn*

## Abstract:

Topological photonics has attracted widespread research attention in the past decade due to its fundamental interest and unique manner in controlling light propagation for advanced applications. Paradigmatic approaches have been proposed to achieve topological phases including topological insulators in a variety of photonic systems. In particular, photonic lattices composed of evanescently coupled waveguide arrays have been employed conveniently to explore and investigate topological physics. In this article, we review our recent work on demonstration of topological phenomena in reconfigurable photonic lattices established by site-to-site cw-laser-writing or multiple-beam optical induction in photorefractive nonlinear crystals. We focus on the study of topological states realized in the celebrated one-dimensional Su-Schrieffer-Heeger lattices, including nonlinear topological edge states and gap solitons, nonlinearity-induced coupling to topological edge states, and nonlinear control of non-Hermitian topological states. In the two-dimensional case, we discuss two typical examples: universal mapping of momentum-space topological singularities through Dirac-like photonic lattices and realization of real-space nontrivial loop states in flatband photonic lattices. Our work illustrates how photorefractive materials can be employed conveniently to build up various synthetic photonic microstructures for topological studies, which may prove relevant and inspiring for exploration of fundamental phenomena in topological systems beyond photonics.

# I. Introduction

Topological photonics has attracted tremendous attention in recent years, drawing concepts and fundamental development from condensed matter physics in which topological phases were first introduced to understand the well-celebrated quantum Hall effect [1,2] and topological insulators [3-5]. In 2005, Haldane and Raghu theoretically proposed the optical analogs of the quantum Hall edge states using photonic crystals with broken time-reversal symmetry [6]. A few years later, Wang et al. further explored the idea and experimentally realized the photonic topological chiral states in two-dimensional (2D) magneto-optical photonic crystals in the microwave regime [7]. However, a drawback of that topological platform is that most optical materials do not possess a large magneto-optical response. In addition, the topological protection of edge state transport depends strongly on the size of the bandgap, so the protection at optical frequencies seems to be impossible due to the small topological gap at these frequencies susceptible to disorder. This has thus hampered the implementation of topological models based on the quantum Hall effect and altogether the advancement of the field in photonics.

Remarkably, several important theoretical proposals and experimental demonstrations of photonic topological insulators emerged in 2013, including the Floquet photonic topological insulators relying on periodic modulation [8], the aperiodic coupled resonators emulating the integer quantum spin Hall [9,10], as well as those based on other mechanisms ranging from topological bianisotropic metamaterials [11] and valley-Hall effects [12], to "network models" of strongly coupled resonators [13,14] and a proposed scheme for crystalline topological insulators [15]. These pioneering endeavors have advanced the study of topological phenomena in photonic systems rapidly, turning the field of topological photonics into one of the frontiers and the most significant branches in topological physics and interdisciplinary sciences. Indeed, there are already several review articles and feature collections dedicated to recent progresses in this field [16-22].

By now, there are several ways to realize topological protected photonic edge states and various photonic topological insulators [16], including for example the photonic topological Anderson insulators [23], nonlinearity-induced topological insulators [24], solitons in photonic topological insulator [25,26] and topological insulators in 3D or even higher synthetic dimensions that normally require complicated fabrication or are not at all possible in real space [27-29]. Towards application, one of the most important technological advancements in the broader field of topological and non-Hermitian photonics is the realization of topological insulator lasers [30-36], which are promising for the future development of low-threshold, robust performance, and single-mode semiconductor lasers.

When it comes to the 1D systems, the existence of topological phases depends on the symmetry of the systems [37,38]. One of the prototypical and most popular 1D models is the so-called Su-Schrieffer-Heeger (SSH) model [39], which is endowed with chiral symmetry and topological modes under nontrivial Zak phase [40]. The first photonic SSH lattice and associated topological states were realized by optical induction using two superimposed periodic patterns to generate the desired dimer chain in a photorefractive crystal [41], as shown in Fig. 1(a). The relative position of the two periodic patterns can be readily tuned to achieve topologically trivial and nontrivial phases of the superlattices. For the nonlinear SSH lattices, it has been shown that nonlinearity can be employed for control of topological zero modes and for inducing spectral tuning and topological phase transition, among other things [42-46]. An example is shown in Fig. 1(b). Normally, the topological modes are located at the two ends (edges) of a nontrivial SSH lattice. However, interfacing two such SSH chains with different Zak phases entails the formation of the topologically protected states at the lattice interface which can be considered as a topological "defect", as realized also in a nanophotonic platform shown in Fig. 1(c) [47]. Interestingly, such a topological defect can serve as a basis for the studies of topologically protected quantum states [48] as well as topological states of parity-time (PT) symmetry in non-Hermitian systems (Fig. 1(d)) [49]. The nontrivial SSH lattices have also been employed to achieve topological lasing, where the lasing mode is manifested by the stationary 1D edge state in the dimer chain of polariton micropillars, as shown in Fig. 1(e) [50], as well as in serval other settings [51-53] including the fabricated silicon microlaser shown in (Fig. 1(f)) [51]. Experimentally, the SSH-type lattices have been vastly implemented in a variety of platforms, including but not limited to photonics and nanophotonics [47,49,54], plasmonics [55,56], and quantum optics [57-59].

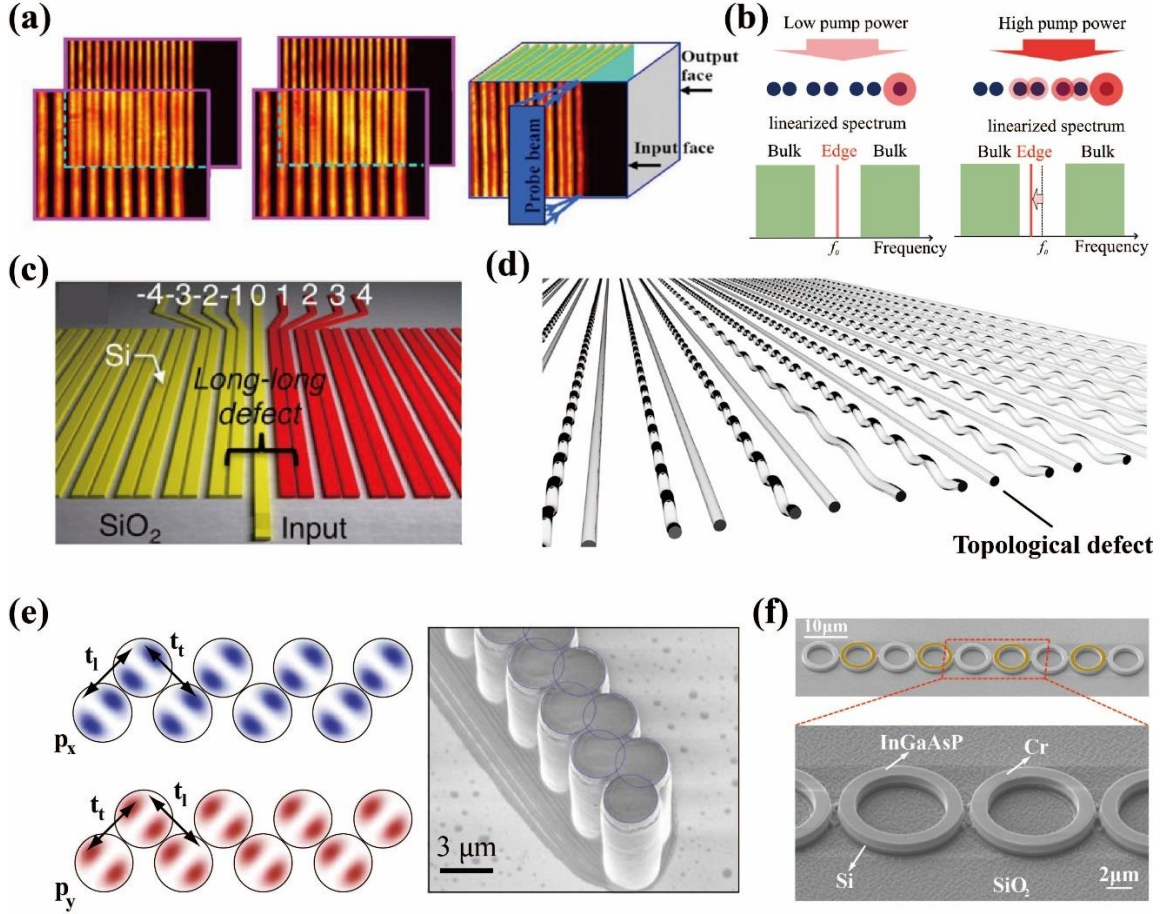

**Fig.1 Photonic SSH lattices realized in various platforms and exemplary applications.** (a) An optically induced SSH dimer chain structure in a photorefractive material [41]. (b) An array of coupled nonlinear resonators in the low and high pump-probe regime [44]. (c) A fabricated nanophotonic SSH lattice with an interface defect in dimerized silicon nanowires [47]. (d) A non-Hermitian PT-symmetric SSH lattice established by fs-laser-writing technique [49]. (e) An SSH lattice constructed via a zigzag chain of coupled micropillars for topological lasing [50]. (f) A fabricated silicon microlaser SSH array [51].

In this paper, we review our recent studies of topological phenomena using photorefractive photonic lattices, with an emphasis on the realization and control of linear and nonlinear topological edge states in the 1D SSH superlattices. In Sec. II, we describe in detail our methods for realization of photonic lattices in photorefractive SBN crystals, including the multi-beam interference-based optical induction and the recently developed continuous-wave (cw) laser site-to-site writing approach. With these two methods, virtually any 1D and 2D photonic lattices can be generated in the bulk of a nonlinear photorefractive crystal. In Sec. III, we present the results of linear and nonlinear topological states in SSH lattices under normal (straight) excitation condition. When a probe beam is tilted to excite the bulk modes of the lattices, nonlinearity induces nontrivial coupling of extended bulk modes into localized topological states, for which a general theoretical framework is developed for interpreting the mode-coupling dynamics in nonlinear topological systems, as discussed in Sec. IV. In Sec. V, a nonlinear non-Hermitian SSH lattice is constructed for the first time to study nonlinearity-induced tuning of PT-symmetry and non-Hermitian topological states. In the 2D lattices, much richer topological phenomena can be expected as compared with that in the 1D domain. For instance, in Sec. VI, we highlight two examples to show what we can do in 2D photorefractive photonic lattices: universal momentum-to-real-space mapping of topological singularities in Dirac-like lattices and nontrivial loop states arising from real space topology in flatband lattices. Finally, in Section VII, we provide a brief summary and outlook for future research along the topological direction. This review is not intended to be overly comprehensive, but rather a concise account of our work on topological photonics by use of photorefractive materials, which may stimulate other groups to start using such materials in the context of topological photonics.

## II. Methods for construction of topological photorefractive lattices

Photonic lattices have been considered as one of the most effective artificial photonic platforms for the study of intriguing discrete phenomena in optics as well as condensed matter and quantum physics due to analogy between the paraxial equation for electromagnetic waves and the Schrödinger equation in quantum mechanics. In other words, the evolution of a wave function with time in quantum systems can be regarded as equivalent to the paraxial electromagnetic waves traveling in space [60,61], which offers a remarkable way to mimic the dynamics of quantum wave functions via classical electromagnetic waves in optics.

In particular, the optical induction technique in photorefractive crystals based on the nonlinear photorefractive effect has been extensively used to study various interesting phenomena including the optical spatial solitons [60,61]. In contrast to the fs-laser-writing technique which permanently changes the refractive index inside the material (typically fused silica glass) [62], optically induced photonic lattices in photorefractive crystals are reconfigurable and non-permanent [63,64]. More specifically, the index change in photorefractive crystals is caused by the intensity-dependent photorefractive effect, so the intensity pattern of the induction beam determines what kind of index potentials (lattices) can be created inside the photorefractive crystals. The lattices are reconfigurable because we can always use a white light source to "wash out" any unwanted index change and re-write with a new intensity distribution. Apart from this advantage and the optical nonlinearity gifted by the photorefractive materials, we mention briefly here what we cannot do or it is difficult to do as compared with the fs-laser-writing technique. Firstly, we cannot optically induce (by multi-beam interference) or point-to-point write (by single-beam launching) long waveguides as limited by the crystal length. The longest crystal we can acquire is only 20 mm long, while the fs-laser-writing glass sample can be several centimeters long. Limited propagation distance is hampering observation of phenomena that become appreciable only after certain distance of beam evolution. Secondly, it is quite a challenge to perform index modulations along the propagation direction in optically induced photonic lattices, although with the latest development we can write sectioned waveguides (to effectively control the waveguide loss) as discussed in a later section. In general, it is virtually impossible to establish $z$-modulated complex lattices such as those obtained with fs-laser-writing – helical waveguides in Floquet topological insulators [8], periodically driven waveguide in anomalous Floquet topological insulators [65], and wiggled waveguide arrays in PT-symmetric lattices [49].

On the other hand, since the index change induced using fs-laser-writing technique relies on high peak intensities to change the material properties, the local index change can only be increased or decreased in a chosen material and become permanent. In this regard, photorefractive crystals have the built-in advantages. Furthermore, the fused silica glass does not have strong optical nonlinearity, but photoreactive materials exhibit high nonlinearity even with a low intensity cw-laser beam. The nonlinearity can be easily tuned by changing the orientation of bias field, which provides an ideal platform to study nonlinear topological phenomena in photonic lattices, apart from early work on optical solitons [61]. Of course, the topological nonlinear edge states and bandgap solitons studied recently [20,24-26] are fundamentally different from those early soliton work. Solitons occur when dispersion or diffraction, which is a linear property of a system, is balanced by the nonlinearity. As linear properties of topologically nontrivial systems are fundamentally different from the topologically trivial ones, this is reflected onto the properties of nonlinear phenomena such as solitons. Recent studies on nonlinear topological photonics are performed in newly discovered topological systems, which were not present decades ago when optical solitons were at the frontier of nonlinear optics.

In the following, we will elaborate two simple methods for creation of photorefractive photonic lattices. One method developed recently is based on site-to-site cw-laser-writing technique (see section IIA, IIB below), which has been used to establish 1D periodic and aperiodic structures [46] as well as 2D complex lattice structures [66-70]. The other method developed earlier based on optical induction [63,71] (see section IIC below), employs periodic or quasi-periodic non-diffracting waves generated by the interference of several plane waves, which is then used to induce various 2D photonic structures used in several experiments with photonic graphene and flatband lattices [72-77].

For the cw-laser-writing technique, we use Gaussian-type beams to write the photonic lattices site by site. If we write from the front entrance to the crystal, as is done for the 2D lattices, the lattice period cannot be too small (say, less than $25\mu m$ in a 10mm-long crystal) due to the limitation of the quasi-nondiffracting zone of the writing beam. Different from the fs-laser-writing technique which is based on scanning the laser beam laterally through the glass sample to write both 1D and 2D lattices, in our case we do not scan the laser beam but rather launch the writing beam longitudinally (from front entrance) or laterally (from sideway) to induce the index change in the photorefractive crystal. We can use either a spot beam (longitudinally along z-direction in Fig. 2a path 1 for 2D lattices) or a stripe beam (laterally along y-direction in Fig. 2a path 2 for 1D lattices) to induce waveguides. The advantages of the sideway writing are addressed below in section IIA, which basically permits the creation of smaller lattice period (thus larger coupling) in a longer crystal sample (say, 20mm, as set by the longest crystal length we have).

*A. CW-laser-writing technique for 1D photonic lattices*

Our experimental method for cw-laser-writing of 1D SSH photonic lattices is illustrated in Fig. 2(a), where the two paths are for probing (path 1) and writing beams (path 2). When a collimated beam (quasi-plane-wave, at wavelength $532nm$ and power $100mW$) illuminates a spatial light modulator (SLM), it is modulated alternatively as the writing and probing beams. For the writing beam, the phase information (the phase to represent a cylindrical lens) encoded onto the SLM turns it into a stripe-type Gaussian beam pattern, for which the waist as well as the position can be readily changed. All the phase information (as the phase mask) is uploaded with the Holoeye Slideshow software and applied onto the SLM one by one with certain repeating frequency.

The writing beam after exiting the SLM is spatially filtered by two lenses (a 4f system) as shown in path 2 in Fig. 2(a). Due to the phase grating added in the predesigned mask, only the first-order diffraction of light can go through this aperture, generating the stripe Gaussian beam. All the zeroth- and other order of light will be blocked by the aperture placed between two lenses. Furthermore, the stripe beam is expanded to laterally cover the entire 20-mm-long photorefractive crystal. To achieve smaller period of the lattices, a set of cylindrical lenses is mounted after the filtering system. Consequently, the stripe beam is compressed into a smaller width before entering the crystal. The waist of the stripe beam is positioned inside the crystal so it remains quasi-nondiffractive while inducing the refractive-index change in the biased crystal. Due to the "memory" effect of the photorefractive crystal, such index change can remain intact more than one hour in the SBN crystal so all index changes can be induced by site-to-site writing and we have enough time for monitoring the probe beam dynamics. To examine the invariance of refractive index potential, we can send a Gaussian probe beam at the input to cover one of the sites and capture the output pattern before and after one hour. The unchanged output pattern indicates that the induced index change remains more or less constant. On the other path, the probe beam is also filtered and collimated through a set of lenses, whose positions are precisely controlled via the SLM. The mask design of the probe beam follows the same procedure but with different beam waist. Linear or nonlinear evolution of the probe beam in the lattice can be easily controlled by switching on and off the bias field. This approach can in principle construct arbitrary 1D lattice structures needed in experiment with spacing between nearest neighbor sites around $15\mu m$. Since the writing beam is launched laterally (illuminating the crystal sideway from its long side), the length of waveguide along z-direction in Fig. 2(a) is therefore determined by the length of crystal. In experiment, we choose a $20mm$-long crystal along z-direction to create the 1D lattices, which provides an effective platform to study the linear and nonlinear topological phenomena in SSH photonic lattices [46], including the trivial lattice (Fig. 2(b1)), the nontrivial lattice with edge (Fig. 2(b2)) and interface (Fig. 2(b3)) defects, and the non-Hermitian SSH lattice (Fig. 2c), as we shall discuss in later sections.

B. *CW laser-writing technique for 2D photonic lattices*

In fact, 2D photonic lattices can also be created when a set of Gaussian-like writing beams is sent into the crystal longitudinally along path 1. Similar with the method mentioned in Ref. [66,67], a cw laser beam (λ=532 nm, with 50 mW power) exiting from the SLM forms a writing beam whose waist position can be precisely controlled by the SLM. The mask design follows the same procedure as we elaborated in above subsection, except the phase distribution for the 2D Gaussian beam is designed and uploaded under this condition. The 4f system in path 1, according to the same method we addressed in previous subsection, filters all other components of light and guarantees a quasi-nondiffracting zone of the writing beam inside the 10mm-long crystal. (Note in this case since the writing beam is launched from the front surface, we cannot use the 20mm-long crystal as otherwise the diffraction of the writing beam cannot guarantee the invariance of the waveguide along *z*-direction). By shifting the relative position between the writing beams via the SLM, the predesigned 2D structures are gradually established due to the crystal "memory" effect. Once the photonic lattice is established, the probe beam also generated in the same path via the SLM is launched into the lattice, which guarantees the perfect alignment between the waveguide channel and the probe beam. In principle, thanks to the flexibility to position the writing beam, any arbitrary 2D waveguides arrays with different boundaries can be established this way, such as 2D SSH-type lattices (Fig. 2d), the super honeycomb [69], and "decorated" honeycomb lattices (HCLs) [78] (Fig. 2e and 2f).

However, limited by the quasi-nondiffracting zone of the writing beam, the period of the lattice cannot be less than $25\mu m$ in the 10mm-long crystal. Fortunately, several phenomena which do not require strong coupling between neighbor sites but only to the next neighbor as in the tight-binding model. For instance, such a method can be used to create the Lieb lattices with bearded edges [67], the Corbino-geometry Kagome lattices [68], the super-honeycomb [69] and the rhombic lattices [76], which are useful for exploration of the flatband localized states and real space topology as discussed in Sec. VI.

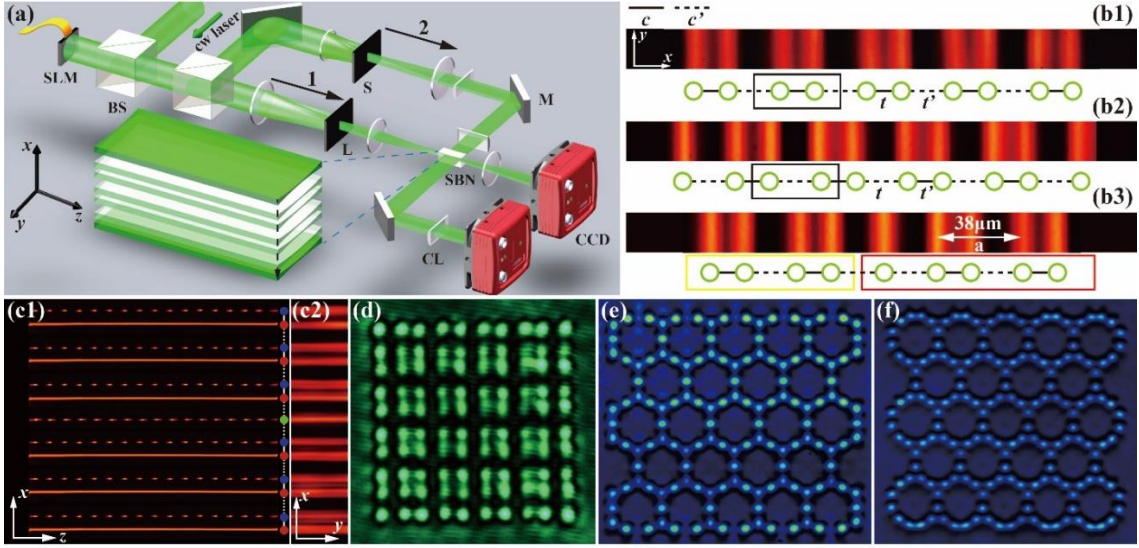

**Fig. 2 Experimental scheme for cw-laser-writing photonic lattices in photorefractive nonlinear crystals.** (a) Experimental setup for writing and probing a 1D lattice. The path 1 is for the lattice-writing beam (ordinarily polarized), and the path 2 is for the probe beam (extraordinarily polarized). Lower inset shows the magnified lattice structure. (b) Illustration of different types of SSH model. $c$ and $c'$ represent strong and weak coupling, black squared zone denotes a unit cell, $t$ and $t'$ are the intra- and inter-cell couplings, respectively [46]. (b1) and (b2) are the trivial, nontrivial SSH structures fabricated in experiment, respectively. (b3) The SSH-type lattice with an interface (long-long) defect formed with one trivial (squared by yellow lines) and one nontrivial (squared by red lines) dimer chains [46]. (c) The passive PT-symmetric SSH lattice with an interface defect, where (c1) is the sideview of the writing beam pattern, with different gaps between sectioned "lossy" waveguides (denoted by blue dots) and the center "neutral" (denoted by green dot on the right side) waveguide. The "gainy" (denoted by red dots) waveguides are all continuous, and (c2) is the written passive PT-symmetric lattice as examined by a broad beam [79]. (d) A 2D SSH-like lattice which can support topological corner states. (e, f) Typical super-honeycomb [69] and decorated honeycomb photonic lattices used in demonstration of flatband states.

## C. Multiple-beam optical induction technique

Due to the diffraction of Gaussian-like writing beams used in cw-laser-writing technique addressed above, the spacing between nearest neighbor sites in 2D structures cannot be reduced to less than $25\mu m$ in a $10mm$ long crystal, which results in the weak coupling between nearest neighbor sites. Therefore, it is not feasible to perform experiments which demand strong coupling or long propagation distance like conical diffractions in 2D lattices [72-75]. Fortunately, multiple-beam optical induction method can be utilized to generate periodic 2D structures, which was first carried out for the observation of 2D optical discrete solitons [63,71]. Comparing with the cw-laser-writing technique, this method can generate 2D periodic structures which can possess longer propagation distance ($20mm$) and smaller nearest neighbor spacing ($9\mu m$) [75], but the disadvantage is that this interference-based method cannot establish periodic lattices with specially desired boundaries or defects, or aperiodic lattices that cannot be formed by multi-beam superposition. For completeness, we briefly describe this method here.

Moreover, nearly the same experiment setup can also be used to construct photorefractive lattices by optical induction, when interfering multiple beams are sent from path 1 as illustrated in Fig. 2(a). Being different from the traditional method in which periodic patterns of the induction beams are generated by an amplitude mask [63,71], here an SLM is used to modulate and control the shape of the induction/writing beams, see in Fig. 2(a). The use of the SLM in this technique provides a more flexible and precise way to control the relative position and periodicity of the writing and probing beams, since then the multiple lattice sites can be written simultaneously instead of site by site. Modulated writing beams from the SLM are spatially filtered by a Fourier filter and eventually form a desired periodic pattern propagating invariantly throughout the photorefractive crystal. With an external electric field applied across the crystal, the desired refractive index patterns (photonic lattices) are formed under the action of nonlinearity. The probe beam can also be modulated by the SLM. Moreover, the relative position between the lattice and the incoming probe beam is a critical parameter in experiment, which is hard to be tuned in traditional induction method without the SLM. Based on this multi-beam induction/writing method, we have created photonic HCLs and Lieb lattices, which are used to demonstrate topological charge conversion as discussed in Sec. VI.

## III. Topological states in 1D SSH-type photonic lattices

In this section, we briefly introduce the topological invariants of the SSH model and the existence of linear topological modes in the nontrivial lattices. Then, we present the results on the study of nonlinear effects on topological interface states in the SSH lattice, which show different behavior under weak and strong nonlinearities.

*A. Linear topological edge and interface states*

The 1D SSH model was originally proposed to describe the charge-transfer dynamics in the long dimer chain of polyenes [39], which by the way is a linear model and has nothing to do with nonlinearity-induced solitons. As illustrated in Fig. 2(b1, b2), a square zone denotes a unit cell, $t$ and $t'$ represent intracell and intercell coupling, respectively. Solid and dashed lines symbolize the strong and weak hopping amplitudes. Under the tight-binding approximation, the SSH model is captured by the Hamiltonian:

$$H = \sum_n t a_n^\dagger b_n + t' a_{n+1} b_n^\dagger + h.c. \tag{1}$$

where $a_n^\dagger (b_n^\dagger)$ are the creation operators on $n$th unit cell. Such a Hamiltonian with chiral symmetry exhibits two bands in the $k$-space, in which corresponding eigenmodes take the form $|\varphi_\pm(k)\rangle$, where $\pm$ indicates the upper and lower bands. The existence of edge states is determined by the winding number which is defined as the Zak phase divided by $\pi$ [40]

$$\mathcal{W}_\pm = \frac{1}{\pi} \oint_{BZ} dk \langle \varphi_\pm | \partial_k | \varphi_\pm \rangle. \tag{2}$$

The value of $\mathcal{W}$ is determined by the relation between intracell and intercell coupling, with $t > t'$ and $t < t'$ corresponding to $\mathcal{W} = 0$ (topologically trivial) and $\mathcal{W} = 1$ (topologically nontrivial), respectively. The most notable difference is that (finite) topologically nontrivial SSH lattice possess localized topological edge modes with eigenvalues in the gap [16], whereas in the topologically trivial phase all modes are extended. For instance, in Fig. 2(b1), the finite SSH lattice is terminated at the weak coupling (i.e., the larger spacing of the dimer), therefore $t > t'$, the winding number is zero (Zak phase 0), resulting in the topological trivial phase without localized edge states. On the other hand, if the lattice is terminated at the strong coupling (i.e., the smaller spacing of the dimer), as in Fig. 2(b2), it will lead to a topological nontrivial system with nonzero winding number (Zak phase $\pi$), which supports a topological edge state [41].

To observe such an edge state, a narrow stripe beam is launched straightly into the edge waveguide, as illustrated in Fig. 3. After 20-mm of linear propagation, it evolves into an edge state in the nontrivial SSH lattice (Fig. 3(b1)) with a characteristic amplitude (and out-of-phase relation) populating only the odd-numbered waveguides counting from the edge (Fig. 3(b2)), which indicates the formation of the topological edge state (the zero-mode residing right in the middle of the band gap) protected by the chiral symmetry. Such an edge state was originally demonstrated in Ref. [41] in 2009, motivated for observation of optical Shockley surface states, but it was realized only later by the community in topological photonics that it in fact represents the first demonstration of 1D topological states in photonics [16,17]. For direct comparison, the corresponding results obtained from the trivial SSH lattice (Fig. 3(a)) show that the input beam couples into a few waveguides nearby the edge which cannot localize in the first waveguide (Fig. 3(a2)). It clearly indicates that the trivial SSH lattice cannot support an edge state under linear conditions.

Moreover, the SSH-type lattice with a center topological defect shown in Fig. 2(b3) can be regarded as the junction of trivial and nontrivial SSH lattices. The interface (here the "long-long" defect or strong-bond termination) SSH-type lattice still possesses chiral symmetry, which guarantees the existence of a topological interface state [47]. When a stripe probe beam is launched straightly into the defect channel as shown in Fig. 3(c1), the output pattern of the probe beam shows strong localization (Fig. 3(c2)), which indicates the existence of topological interface state.

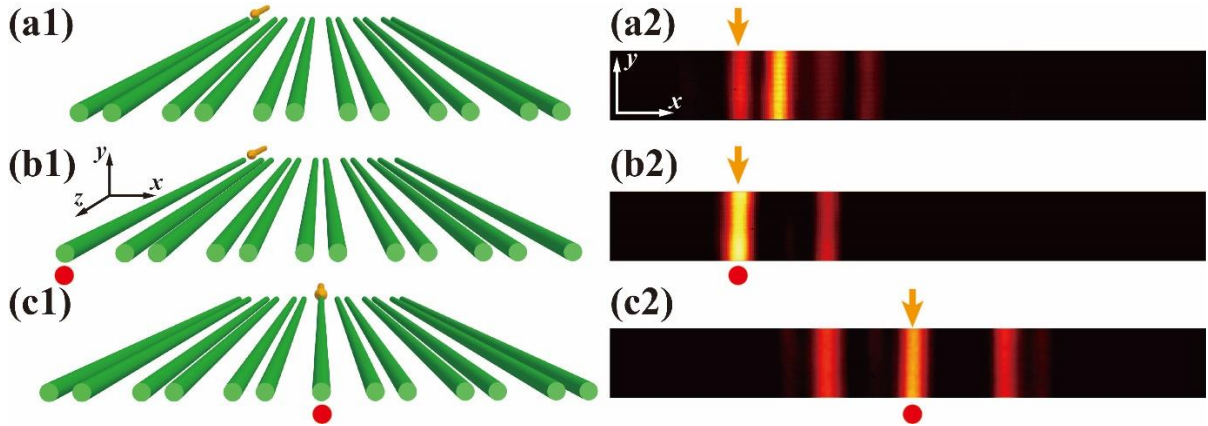

**Fig. 3 Experiment demonstration of linear topological states in 1D SSH lattices.** (a) to (c) show the illustration of (a1) trivial and (b1, c1) nontrivial SSH lattices and the corresponding output intensity patterns of a probe beam, showing (a2) discrete diffraction and characteristic topological (b2) edge and (c2) interface states. The excitation position is denoted by the arrows, and the red dots mark the topological defect channel. Notice that there is no light field in the lattice site nearest to the defect in (b2) and (c2) [46].

### B. Nonlinear topological interface states

In this subsection, we present the key results of nonlinear topological interface states under both self-focusing and self-defocusing nonlinearity. The experiment results are summarized in Fig. 4. The SSH lattice structure is the same as the one shown in Fig. 2(b3). When the nonlinearity is turned on by applying an electric field ($+8.8 \times 10^5 V/m$ for self-focusing, and $-8 \times 10^5 V/m$ for self-defocusing) and the probe beam itself has low power, the output patterns still resemble the linear topological state (Fig.4(a2, a4)), indicating the formation of a topological gap soliton under weak nonlinearity. However, the output patterns change drastically under strong nonlinearity by increasing the power of the probe beam. Specifically, under strong self-focusing nonlinearity, the probe beam is confined into the center defect (Fig. 4(a1)), corresponding to the formation of a soliton with propagation constant residing in the semi-infinite gap [42,46]. On the other hand, the output pattern becomes strongly delocalized, spreading into the bulk of the lattice under the action of a high self-defocusing nonlinearity, as shown in Fig.4(a5). These results show that the topological states in SSH lattices can sustain only weak nonlinearity, but they cannot maintain the characteristics of the topological states under strong nonlinearity, as corroborated in theory [80].

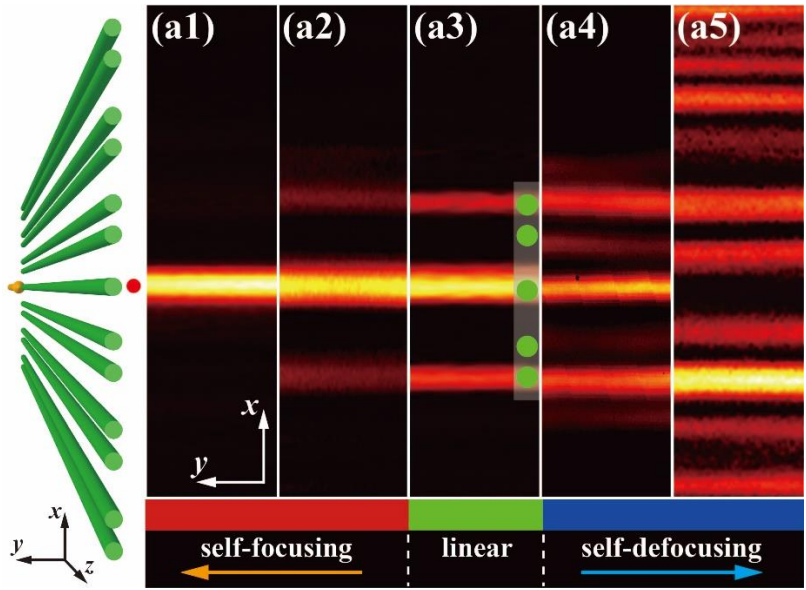

**Fig. 4 Experiment results showing nonlinear effects on a topological interface state.** Left panel: illustration of single-channel straight excitation at the defect channel (red dot). (a) Output transverse intensity profiles of the probe beam under (a1) strong, (a2) weak self-focusing nonlinearity, and (a3) linear propagation, and (a4) weak and (a5) strong self-defocusing nonlinearity. Red, green and blue bars in the bottom of (a) denote the system under self-focusing, linear and self-defocusing conditions, respectively [80].

## IV. Nonlinearity-induced coupling of light into topological states

In this Section, we present the key results from our recent studies of nonlinearity-induced coupling of light into a topological defect [46]. Different from that presented in Sec. III, in which the topological modes are excited by the probe beams initially launched straightly along the defect channel, here a tilted broad beam travels across several lattice sites towards the defect. In this recent work, we have been able to show nontrivial coupling of light into the defect channel due to the nonlinearity-induced mode coupling between bulk modes and topological states, as explained by our theoretical analysis based on nonlinear mode-coupling dynamics in topological systems [46].

Both trivial and nontrivial SSH lattices are established using the cw-laser-writing technique in our experiment to study the nonlinear mode coupling. In a trivial lattice, a tilted broad beam ($k_x = 1.4\pi/a$, and the lattice constant $a = 38\mu m$) could couple into the edge waveguides (Fig. 5(a1)) due to the lack of topological protection but escape from the edge under nonlinear condition (Fig. 5(a2)). The nonlinear sideview in Fig. 5(a3) indicates that the energy of the initial beam gradually dissipates into the bulk in this trivial system. On the contrary, when the same tilted broad beam is sent into the nontrivial lattice, due to the topological protection of the nontrivial SSH lattice (which arises from the chiral symmetry), no light can couple into the edge under linear propagation (Fig. 5(b1)). Fig. 5(b2) shows the nonlinear induced coupling of light into topological states. In this case, nonlinearity locally breaks the chiral symmetry and traps a portion of light into the topological protected edge waveguide in the nontrivial system (Fig. 5(b3)). This is the key results delivered in this work.

Moreover, we have explored the interaction of two tilted beams from opposite directions with respect to the defect. The SSH lattice with a topological defect in the middle can also be established as shown in Fig. 2(b3). The phase relations between the two probe beams determine the mode coupling between bulk and defect mode in nonlinear regime, where mode coupling gets enhanced under in-phase excitation (Fig. 6(a2)) but suppressed under out-of-phase excitation (Fig. 6(b2)).

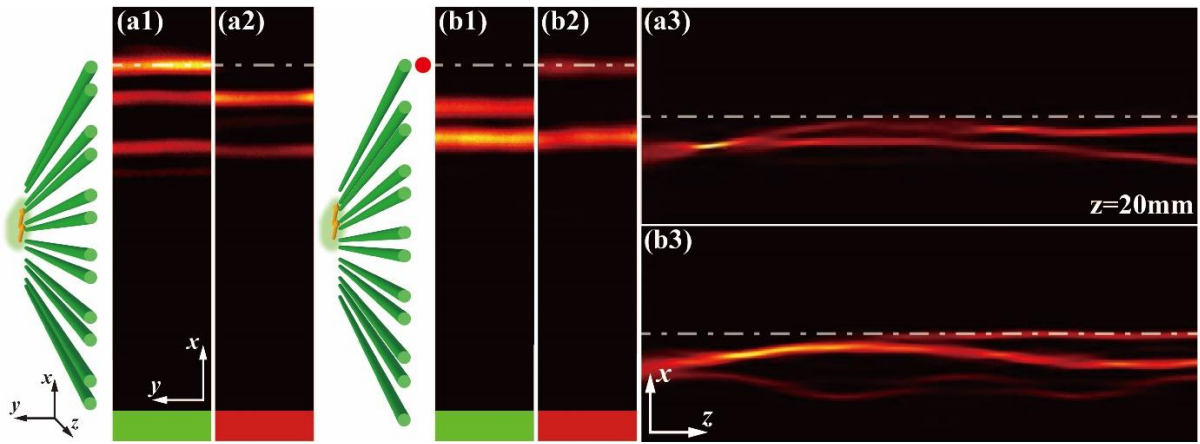

**Fig. 5 Nonlinearity-induced coupling into a topological edge state in nontrivial SSH lattices.** The two illustrations show tilted excitation from bulk to edge for the trivial (left) and nontrivial (right) lattices, where the
15

red dot marks the position of the nontrivial edge. Results in (a) are for trivial lattice, where (a1) and (a2) show the experimental outputs of the probe beam (initially tilted at $k_x = -1.4\pi/a$) in the linear and nonlinear cases, respectively, and (a3) is side-view propagation from corresponding simulation (up to the crystal length of 20 mm) under nonlinear excitation. Results in (b) have the same layout as in (a) except that they are from the nontrivial case. The white dashed-dotted line marks the position of the edge channel in the SSH lattices in all figures [46].

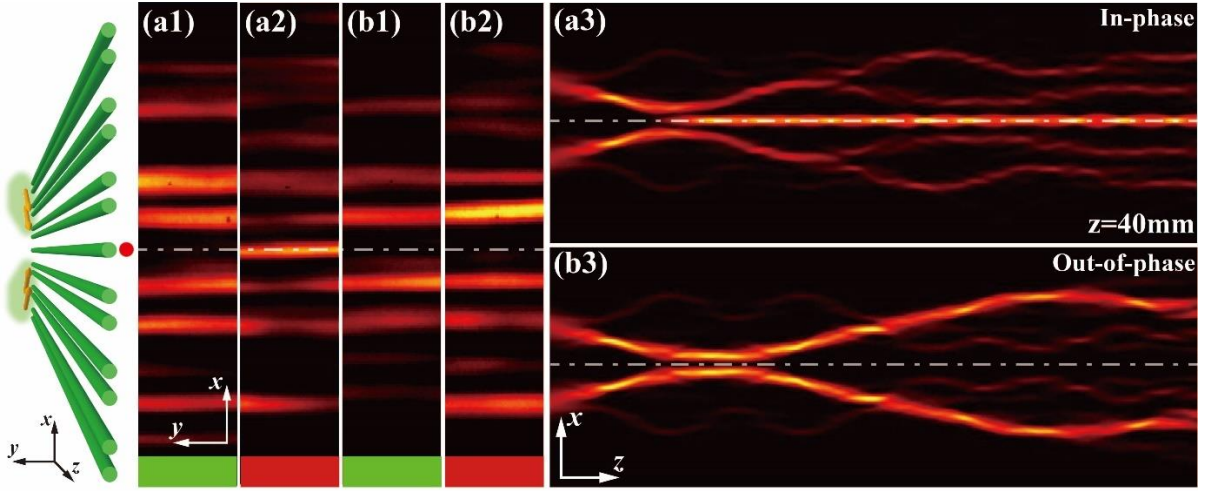

**Fig. 6 Nonlinearity-induced coupling and "escaping" of light at a topological interface defect.** The illustration on the left panel shows two-beam tilted excitations of the topological interface defect (marked by red dot) from opposite directions. (a1) and (a2) show the experimental outputs of two tilted in-phase beams ($k_x = \pm 1.4\pi/a$) under (a1) linear and (a2) nonlinear excitation conditions, and corresponding simulation results are displaced in (a3), showing a sideview of the beam dynamics (up to a length of 40 mm) under nonlinear excitation. Results in (b1–b3) have the same layout as in (a1–a3) except that the defect is excited with two tilted out-of-phase beams, showing "escaping" of light from the defect in nonlinear condition. The white dashed-dotted line marks the position of the nontrivial interface defect channel in the SSH lattice in all figures [46].

The underlying mechanism is analyzed on the basis of nonlinear coupled mode theory [46], which explains well the experimental observations in Fig. 5 and Fig. 6. Take the results in Fig. 5(b) as a specific example, the theory provides a basic understanding of the nonlinearity-induced mode coupling in the nonlinear SSH model. Briefly, the mode-coupling dynamics undergoes three different stages. In the first stage, the probe beam, although undergoing nonlinear propagation, cannot excite the linear topological edge mode. As the beam approaches the defect, in the middle stage, it strongly perturbs the local structure of the lattice at the edge, so topological edge mode does not exist. Eventually, in the third stage, part of the probe beam evolves into a nonlinearly localized mode, and the rest of the light escapes into the bulk..

In fact, our theory shows that the nonlinear edge mode has inherited properties from the topological edge state [46]. To be more specific, our study addressed the distinction between *inherited* and *emergent* nonlinear topological phenomena. The inherited nonlinear topological states are similar to the modes of the underlying linear system, whose corresponding linear mode will be affected by nonlinearity but without closing the gap or changing the topological invariants. Such similarities can be examined by the mode overlapping between the nonlinear state and its corresponding linear mode. On the other hand, emergent nonlinear topological phenomena arise from the linear trivial systems, where nonlinear dynamics change the topological invariants and turn the system into a topologically nontrivial regime. Therefore, the nonlinear edge mode shown in Fig. 5 and Fig. 6 features a topological edge state fully inherited from the underlying linear system. These features exemplify the interplay of topology and nonlinearity in topologically nontrivial systems, leaving many intriguing questions yet to be explored [46].

## V. Nonlinear control of non-Hermitian topological states

In the past decade, the study of PT-symmetry in optics has offered a novel strategy for controlling light by manipulating the relation between gain and loss in non-Hermitian systems, leading to rapid advancement of the field of non-Hermitian optics [81-86]. However, the existence of PT-symmetric topological edge mode was initially considered impossible, since the existence of topologically protected edge states tends to break the PT-symmetry in non-Hermitian systems which results in complex eigenvalues [87]. Recently, linear PT-symmetric topological states with real eigenvalues have been observed in non-Hermitian SSH-type lattices fabricated by fs-laser-writing (see Fig. 1(d)) [49]. In this Section, we present the key results from our recent study based on nonlinear non-Hermitian SSH lattices, demonstrating for the first time nonlinear tuning of the PT symmetry and topological states in such intriguing systems [79].

*A. The non-Hermitian PT-symmetric SSH model*

The non-Hermitian PT-symmetric SSH model is sketched in the left panel of Fig. 7, where red, green and blue rods denote the "gainy", "neutral" and "lossy" waveguides. The Hamiltonian of such a PT-symmetric SSH model can be written as following, under tight-binding approximation:

$$\mathcal{H} = c \sum_{n \in N_G} a^\dagger_{n-1} a_n + c' \sum_{n \in N_G} a^\dagger_{n+1} a_n + c' a^\dagger_1 a_0 + h.c. + \sum_{n \in N_G} \left(\beta^* a^\dagger_n a_n + \beta a^\dagger_{n-1} a_{n-1}\right) + \beta_0 a^\dagger_0 a_0$$

$$N_G = \{-1, -3, -5, \cdots\} \cup \{2, 4, 6, \cdots\} \tag{6}$$

where $\beta = \alpha + i\gamma$, $\alpha$ is the real part of a waveguide potential, and $\gamma$ is the imaginary part representing the gain or loss. $c$ and $c'$ are the strong and weak couplings (illustrated by smaller and larger distances between adjacent waveguides in Fig. 7). $a^\dagger_n$ is the creation operator on the $n$th site. The set of integers $N_G$ corresponds to all the "gainy" sites marked as red channels in Fig. 7.

Thus far, optical nonlinearity is normally regarded as an effect which changes the refractive index and can lead to the phase transition in PT-symmetric system as previous works stated [88]. However, counter-intuitively, as we have demonstrated quite recently [79], even though the photorefractive nonlinearity used in our experiment induces a change only in the real part of refractive index, it can equivalently affect both the real and imaginary parts of an induced waveguide potential. Therefore, once the passive PT-symmetric SSH lattice is established in experiment, we can employ optical nonlinearity to change the imaginary part of center waveguide potential and thus turn the whole lattice into a non-PT symmetric system. As such, the eigenmode profiles become asymmetric with respect to the center defect, which can serve as a signature of phase transition between PT and non-PT systems [79].

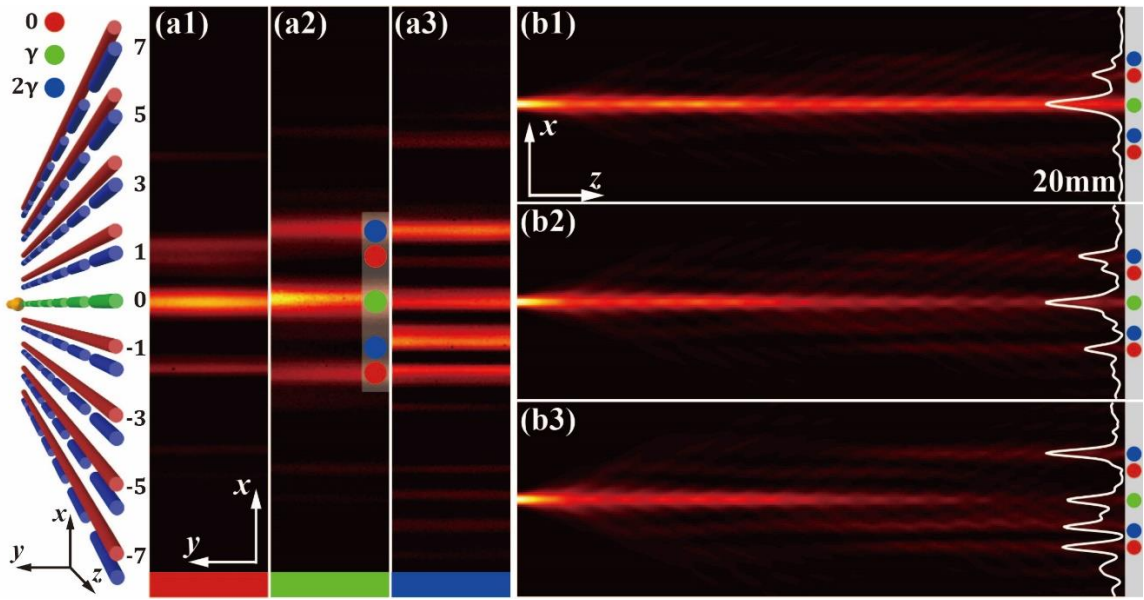

**Fig. 7 Observation of single-channel nonlinear control of PT-symmetry and topological interface state.** Left panel illustrates the non-Hermitian SSH lattice as shown in Fig. 1(d), which consists of continuous and sectioned waveguides. (a1-a3) Experimental results showing output transverse patterns of a probe beam launched into the center defect channel. (b1-b3) Simulation results showing sideview propagation corresponding to (a1-a3). The SSH lattice is initially under passive PT-symmetry, which supports a linear PT-symmetric topological state (a2, b2). It turns into a non-PT "gainy" system under the action of a self-focusing nonlinearity, so the mid-gap state becomes asymmetric as more energy goes to the nearby "gainy" waveguides (a1, b1). The situation for transition to a non-PT "lossy" system under a self-defocusing nonlinearity is shown in (a3, b3), in which more energy goes to the nearby "lossy" waveguides. This can be seen more clearly from superimposed intensity profiles at $z = 20mm$ (white lines) in (b1-b3). [79]

*B. Nonlinear control of non-Hermitian topological edge states.*

To demonstrate single-channel nonlinear tuning of PT phase transition in an experimental system without actual gain, a global loss is often added into the corresponding active linear non-Hermitian PT-symmetric system as an offset, which leads to a passive PT-symmetric system [49,81]. Apart from fs-laser-written wiggled or scattered waveguides to introduce loss [49,89,90], lossy metal stripes were mounted on top of semiconductor or silicon waveguides to introduce loss [81,91]. In our work, the non-Hermitian SSH lattices are established by employing direct cw-laser-writing technique [46,67] where the loss is introduced by writing sectioned waveguides as illustrated in Fig. 7. The gap size introduced between sections controls the amount of loss in each waveguide to construct "lossy" and "neutral" waveguides [79].

After a passive PT-symmetric SSH lattice is constructed, a stripe beam is sent into the center "neutral" waveguide channel. When the probe beam linearly travels through the lattice, a symmetric topological interface state is observed as shown in Fig. 7(a2), indicating that the non-Hermitian lattice in this condition satisfies the PT-symmetry [49]. Since the SSH lattice is fabricated inside the photorefractive crystal, self-focusing and -defocusing nonlinearities can be easily achieved by applying positive or negative bias field along the crystalline c-axis of the photorefractive crystal [60,61]. With the buildup of defocusing nonlinearity, the probe beam escapes more from the center defect, and the intensity distribution at the output is highly asymmetric as shown in Fig. 7(a3), indicating that the self-defocusing nonlinearity turns the system from passive PT-symmetric to non-PT "lossy" regime. On the contrary, under the action of self-focusing nonlinearity, the probe beam distributes more into the nearby "gainy" waveguide as shown in Fig. 7(a1), indicating that the lattice is now in non-PT "gainy" regime. Corresponding simulation results in Fig. 7(b) show clearly the transition between symmetric to asymmetric interface states due to the action of nonlinearity. On the other hand, because the sign of nonlinearity in photorefractive crystals can be easily reversed, the transition between PT and non-PT regimes can be reversed also by nonlinearity, leading to destruction or restoring of PT-symmetric topological states [79]. This opens many possibilities for the study of nonlinear non-Hermitian topological systems, including for example nonlinearity-induced birth and death of zero modes and exceptional points in non-Hermitian systems.

## VI. Topological phenomena in 2D photorefractive photonic lattices

In previous sections, we reviewed mainly 1D photorefractive structures where topological edge states are determined by the trivial/nontrivial winding number. However, apart from the topological edge/interface states, there are many other phenomena that also have topological origin, such as pseudospin conversion in photonic lattices caused by the Berry phase encircling the Dirac-like cones [75] and non-contractible loop states in photonic flatband lattices such as the Lieb [67] and Kagome lattices [68] arising from real space topology. In fact, over the years, we have used photorefractive Dirac-like photonic lattices to study novel phenomena that would otherwise be inaccessible in natural 2D materials, including pseudospin excitation and valley vortex states, valley Bloch oscillations, and unusual flatband localized states. In this last section, we discuss briefly two examples where photorefractive photonic lattices have been employed as a versatile platform to study topological phenomena mediated by the Dirac cone and the flat band with singular band-touching.

*A. Topological singularity mapping in Dirac-like photonic lattices*

Topological properties of lattices are typically presented in momentum space, described by the Berry phase [92]. Berry phase in Bloch bands is defined as $\gamma = i\oint d\boldsymbol{k} \langle \varphi_{n,\boldsymbol{k}} | \boldsymbol{\nabla}_{\boldsymbol{k}} | \varphi_{n,\boldsymbol{k}} \rangle$, where $|\varphi_{n,\boldsymbol{k}}\rangle$ describes the Bloch state of the $n$-th band, and the integral is performed over a loop in $k$-space. As such, Berry phase describes the geometrical properties of $|\varphi_{n,\boldsymbol{k}}\rangle$ varying as a function of momentum vector $\boldsymbol{k}$. When one integrates over a path encircling a Dirac point in an HCL, a non-trivial Berry phase of $\pi$ is found; the value of $\gamma/2\pi$ is encoded by the valley Chern number (similar to the winding number described earlier for the SSH model), which is for example used to describe the valley Hall effect [12,93-95]. Recently [75], we have found a universal mapping of topological singularity (Dirac point) from momentum to real space, which shows that, except the valley Hall effect, topological singularity in real space can also be seen as a manifestation of valley Chern number.

Let us consider two specific systems which have Dirac-like cones in momentum space, i.e. the HCL (pseudospin 1/2) and the Lieb lattice (pseudospin 1). Generally, the wave dynamics around a Dirac point in such systems is governed by the effective Hamiltonian

$$H = \kappa \mathbf{S} \cdot \mathbf{k} \qquad (7)$$

where $\mathbf{S}$ denotes the pseudospin angular momentum operator, $\mathbf{k}$ is the displacement of the transverse wavevectors with respect to the Dirac point, and $\kappa$ depends on the properties of the system. The eigenmodes of the Hamiltonian satisfy $H\psi_{n,\mathbf{k}} = \beta_{n,\mathbf{k}}\psi_{n,\mathbf{k}}$. An optimal topological charge conversion occurs with the rule $l \to l + 2s$, where $l$ denotes the orbital angular momentum of the initial input beam, $s$ is the initially excited pseudospin eigenstate, as illustrated in Fig. 8(a, d) [75]. The HCL hosts a two-band touching at the conical intersection, corresponding to pseudospin modes $s = \pm 1/2$ (Fig. 8(a)), whereas the Lieb lattice hosts a three-band touching corresponding to pseudospin modes $s = \pm 1, 0$ (Fig. 8(d)) [96].

In our experiments, both HCL (Fig. 8(b)) and Lieb (Fig. 8(e)) lattices are fabricated using the optical induction method introduced in Sec. II. In Fig. 8(c1), a vortex with a topological charge $l = 1$ is sent into the HCL to excite the pseudospin state $s = 1/2$, and the output turns to a vortex with topological charge $l = 2$. To observe the topological charges experimentally, a tilted plane wave is launched to obtain the interferogram with the output beam [72-75]. Due to the phase singularity in a vortex beam with topological charges, the interference pattern between the output beam and the plane-wave reference beam will generate fringe bifurcations, which reveal the position as well as the value of the topological charges as shown in Fig. 8(c2). In the Lieb lattice, however, the optimal excitation ($l = 1, s = 1$) leads to generation of output vortex with charge $l = 3$. These observed results show clearly that the optimal topological conversions satisfy the rule $l \to l + 2s$ as shown in Fig. 8(c2) and Fig. 8(f2). Theoretically, such a conversion has a topological origin which is related to the Berry phase, and can be rewritten with a more general form $l \to l \pm w$, where $w$ is the winding of nontrivial Berry phase around Dirac-like cones. Such a mapping is universal and can also be generalized to 3D topological singularities such as Weyl points [75].

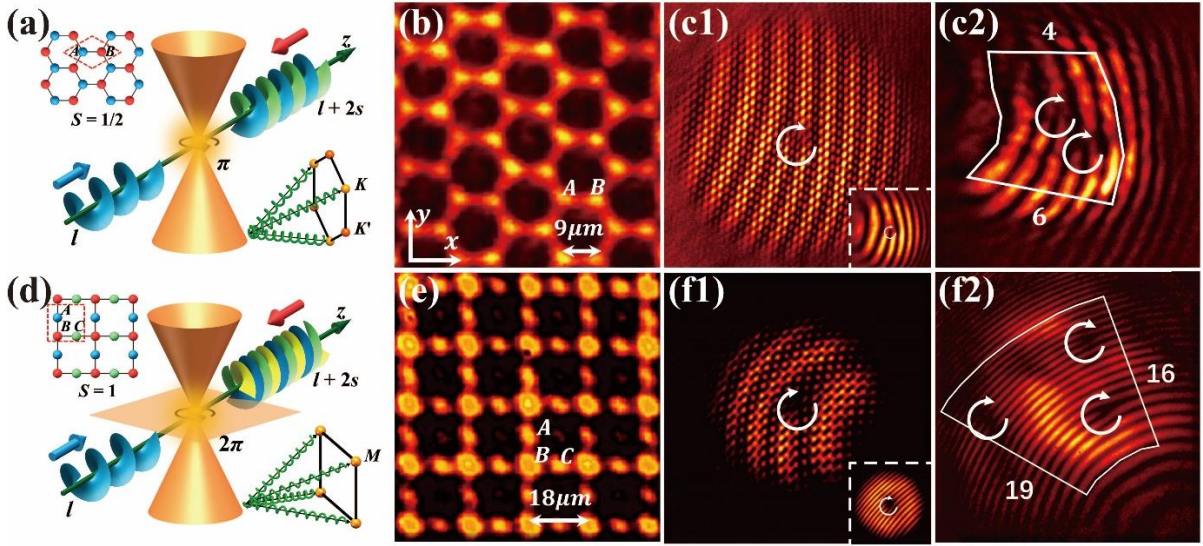

**Fig. 8 Momentum-to-real-space mapping of topological singularities in Dirac-like photonic lattices.** (a) A pseudospin-1/2 HCL with two sublattices *A* and *B* is excited with three vortex beams, each with a topological charge $l$. These vortex beams excite modes around the conical intersections at the corners of the Brillouin zone (lower right inset). The arrow circulating around the Dirac cone illustrates the winding of the Berry phase ($\pi$ in HCL). (b) An experimentally established HCL by optical induction. (c) Excitation of $s = 1/2$ pseudospin state with vortex beams of initial topological charge $l = 1$ (c1), leading to topological charge conversion from 1 to 2 (c2). Bottom panels (d-f) are for the Lieb lattice, which has three sites (*A, B, C*) per unit cell, showing topological charge conversion from 1 to 3 under the excitation condition $s = 1$, $l = 1$. White curved arrows mark the position and helicity of the vortices. [75]

*B. Unconventional flatband states arising from real space topology*

In 2015, two independent groups demonstrated the compact localized states (CLSs) as the eigenstates of the flatband lattices [97,98], which shows clearly that waves can stay localized in the continuum without the presence of any defect or nonlinearity. Those experiments were performed in photonic Lieb lattices, which have one flatband touching the dispersive bands. These early demonstrations stimulated a great deal of interest in flatband study [66,99], especially when the interplay between the concepts of topology and flatband is considered. In fact, according to the early theoretical work in Ref. [100], band-touching between flatband and dispersive bands brings about unconventional flatband eigenstates that arise from the real space topology. In such flatband systems, the conventional CLSs are not complete, with the missing states manifested as non-contractible loop states (NLSs) wrapping around the entire lattices (or the torus representing an infinite lattice system). Such flatband states are topologically distinct from the conventional CLSs, as they cannot be continuously deformed into the CLSs in the torus geometry as shown in Fig. 9(a1). Theoretically, NLSs exists only in an infinite system which is not feasible in experiment. We have shown two different ways to demonstrate such NLSs in finite lattices with different geometries.

In the first case, we have used a finite photonic Lieb lattice with bearded edges (depicted in Fig. 9(a2)) to observe the flatband line states. When flatband lattices are characterized as bipartite systems such as the Lieb [67] and super-honeycomb lattices [69], they support the compact localized line states under appropriate boundary terminations. Such line states can be regarded as the manifestation of the NLSs in finite systems, as they cannot be obtained by linear superposition of the conventional CLSs. To observe the line states experimentally, a Lieb lattice is established in the photorefractive material (Fig. 9(b)). A line-shaped beam with appropriate phase (Fig. 9(c1)) is sent into the lattice with bearded edge. It evolves into a flatband line state with its overall intensity pattern well maintained (Fig. 9(c2)), even after long propagation according to corresponding simulations (Fig. 9(c3)), which proves the existence of the line states in the Lieb lattices.

In the second case, direct observation of the NLSs is achieved by using a Corbino-geometry to mimic the infinite system in one direction. Such a Corbino geometry is obtained by warping a Kagome lattice ribbon into a ring, defining a 2D system confined between two concentric circles (Fig. 9(a3)) [68]. Corresponding lattice has been fabricated in experiment (Fig. 9(d)) using the site-by-site writing technique introduced earlier. Subsequently, a ring-shaped necklace pattern with out-of-phase condition (inset in Fig. 9(e)) is launched into the Corbino-shaped Kagome lattice to excite the NLS sketched in Fig. 9(a3). Corresponding experimental results are shown in Fig. 9(e), where the necklace beam remains intact after 10mm of propagation. Results in Fig. 9(d, e) represent a direct demonstration of the NLSs which are originally proposed as the nontrivial flatband eigenstates for an infinite system.

The study of artificial flatband lattices will deepen our understanding of fundamental concepts and unveil many intriguing phenomena in the interdisciplinary areas such as PT-symmetric flatbands [89], flatband induced type-III and tilted Dirac cones [101] and localized states in 2D Moiré lattices [102].

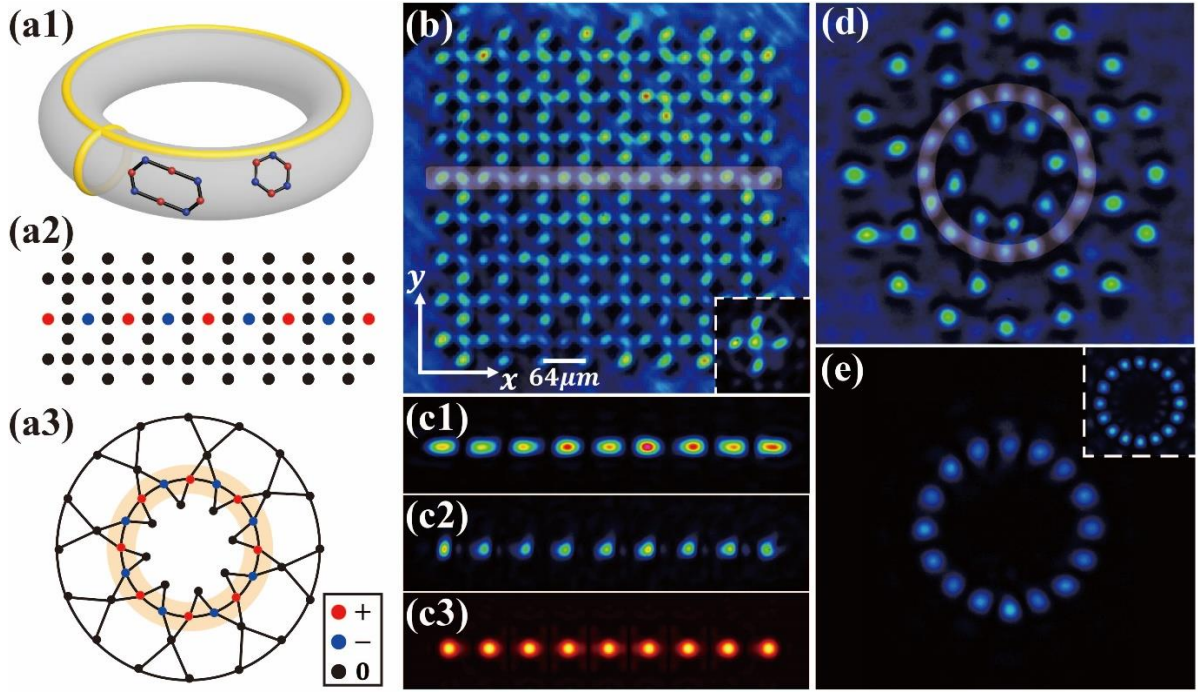

**Fig. 9. Nontrivial flatband line (loop) states in laser-written Lieb (Corbino-Kagome) photonic lattices.** (a1) Two NLSs (yellow) as well as CLSs (black) are illustrated in a torus mimicking a 2D infinite lattice. (a2) Schematic shows an unconventional line state in a finite Lieb lattice terminated with bearded edges. Sites with zero amplitudes are denoted by black dots, and those with nonzero amplitudes of opposite phase by red and blue dots [67]. (a3) Schematic shows a Corbino-shaped Kagome lattice, where the NLS is illustrated by the orange circle [68]. (b) A photonic Lieb lattice established with the cw-laser writing technique, and the shadow marks the position of input beams [67]. (c1, c2) Transverse intensity patterns of an out-of-phase probe beam at (c1) input and (c2) output through the lattice written in a $10mm$ long crystal. (c3) Simulation results showing the line state remains intact after propagating a long distance of $40mm$ through the lattice [67]. (d) A Corbino-shaped Kagome lattice written in the crystal [68]. (e) An NLS observed in experiment under the excitation by an out-of-phase necklace-like probe beam as shown in the inset [68].

## VII. Summary and outlook

We have briefly reviewed recent experimental progress of realization of novel topological states in 1D and 2D photorefractive photonic lattices. Utilizing the optical induction technique, a variety of topological systems can be established in the photorefractive crystals, such as the SSH-type lattices, and the HCL, Lieb and Kagome lattices [41,42,46,66-70,72-77,79,80]. Taking advantage of the nonlinearity of the photorefractive crystals, the interplay between nonlinearity and topological systems can be studied in a convenient photonic platform, which provides an ideal testbed for nonlinear topological photonics in SSH photonic lattices. More interestingly, the nonlinearity-induced tuning of PT symmetry and topological edge states can be investigated in the non-Hermitian SSH lattices. The research on interaction between nonlinearity, topology and non-Hermiticity will lead to a new scheme for control of complex systems and contribute to the development of advanced photonic devices. In addition, a few typical examples of 2D topological phenomena arising from momentum-space topological singularity and real-space topological structures are also presented.

Before closing, we want to briefly discuss several potential directions for research into topological phenomena with photorefractive photonic lattices, such as topological states in trimer-like lattices, fractionalized corner states and related nonlinear processes. Trimer lattices can hold a variety of topological states by controlling the coupling between sites in the system [103]. When intracell couplings are not equal, which breaks the chiral symmetry of the system, the winding number is no longer integer [37,38]. However, such systems without chiral symmetry still support edge states immune to perturbation. Such topological edge states certainly merit further investigation. Secondly, recently established paradigm of higher-order topological states has shown the existence of localized corner states which have robust properties [104-109]. The observation of these novel states can be realized using fs-laser-writing technique, which may also be accessible via optical induction or cw-laser-writing in nonlinear photorefractive materials. Moreover, structures written in photorefractive crystals provide a platform to explore nonlinear effects in topological systems. For example, corner states could be intertwined with edge modes due to the nonlinear self-focusing and -defocusing effects.

In summary, compared with various platforms widely used in optics, the structures fabricated in photorefractive crystals via optical induction and weak-light cw-laser-writing method have great advantages for many different research directions such as in 1D topological structures, flatband structures, and non-Hermitian topological systems, especially when nonlinearity comes to play the role. It is also advantaging to use the photorefractive lattices as a tunable platform since they can be used for investigation of nontrivial fundamental physics and they can be readily implemented in the lab even by less-experienced undergraduate students. Importantly, the investigation on topological physics in such an optical setting can deepen our understanding of relevant fundamental concepts, and in particular, in the largely under-explored areas such as nonlinear effects in topological flatbands, nonlinear higher-order topological states, nonlinearity-induced topological phase transition, and nonlinearity-induced tuning of PT symmetry and exceptional points in non-Hermitian systems. Furthermore, the concepts and ideas developed with the simple optical setting may eventually contribute to the development of light manipulation and novel technological applications, for instance, in nonlinear control of mode coupling and PT transition served as optical switches, robust topological charge generation and conversion, and flatband-assisted image transmission in waveguide arrays, to name just a few.

**Acknowledgment:**

We thank D. Leykam, K. Makris, A. Szameit, J.W. Rhim, J.Yang and all our collaborators on related work.

**Funding:** This research is supported by the National Key R&D Program of China under Grant No. 2017YFA0303800, the National Natural Science Foundation (11922408, 91750204, 11674180), PCSIRT, and the 111 Project (No. B07013) in China. H.B. acknowledge support in part by the Croatian Science Foundation Grant No. IP-2016-06-5885 SynthMagIA, and the QuantiXLie Center of Excellence, a project co-financed by the Croatian Government and European Union through the European Regional Development Fund - the Competitiveness and Cohesion Operational Programme (Grant KK.01.1.1.01.0004).

**Disclosures.** The authors declare no conflicts of interest.

**Authors' contributions**:   All authors contributed to this work.